# Constraints on General Relativity Geodesics by a Covariant Geometric Uncertainty Principle

**David Escors [1]\* and Grazyna Kochan [2]**

[1]  Fundacion Miguel Servet. Public University of Navarre, UPNA; descorsm@navarra.es
[2]  Fundacion Miguel Servet. Public University of Navarre, UPNA; grazyna.kochan@navarra.es
\*  Correspondence: descorsm@navarra.es

**Abstract:** The classical uncertainty principle inequalities were imposed over the general relativity geodesic equation as a mathematical constraint. In this way, the uncertainty principle was reformulated in terms of proper space–time length element, Planck length and a geodesic-derived scalar, leading to a geometric expression for the uncertainty principle (GeUP). This re-formulation confirmed the need for a minimum length of space–time line element in the geodesic, which depended on a Lorentz-covariant geodesic-derived scalar. In agreement with quantum gravity theories, GeUP imposed a perturbation over the background Minkowski metric unrelated to classical gravity. When applied to the Schwarzschild metric, a geodesic exclusion zone was found around the singularity where uncertainty in space-time diverged to infinity.

**Keywords:** general relativity; uncertainty principle; geodesics; black hole singularity; zero-point energy; quantum gravity; Planck star

## 1. Introduction

General Relativity (GR) describes gravitation as a dynamical space–time geometry in a pseudo-Riemannian manifold shaped by energy-momentum densities [1]. Its mathematical framework is consistent and valid in any reference frame. However, GR is largely incompatible with quantum mechanics. For example, GR world-lines for particles are defined with infinite precision [1], while this is not allowed in quantum mechanics. Measuring position in quantum mechanics introduces uncertainty in momentum and vice versa [2]. The momentum/position uncertainty originally proposed by Heisenberg is considered a fundamental principle in nature [3], which is behind many quantum phenomena [4,5]. Other major difficulties in reconciling general relativity with quantum mechanics are the non-renormalizability of GR when formulated as a quantum field theory [6] and the quantum-mechanical violation of the weak equivalence principle [7].

In quantum gravity theories, a limit on the length of the space–time line element is imposed when energy fluctuations alter the space–time metric [8]. These quantum fluctuations may constitute part of the source of a gravitational background state that imposes such a length limit [8]. However, the concept of such a background state presents difficulties because classical GR is a background-independent theory, in which the space–time metric is the dynamical variable [9]. String theory leads naturally to expressions dependent on a fundamental length [10], from which a generalized uncertainty principle (GUP) associated with Planck length is derived [8,11]:

$$\Delta x \Delta p \geq \frac{\hbar}{2} + \frac{k}{c^3} G \Delta p^2. \tag{1}$$

Here, $\Delta \boldsymbol{p}$ represents the change in magnitude of momentum parametrized by coordinate time; $\Delta \boldsymbol{x}$ is the change in magnitude of the position vector; $\hbar$ is the reduced Plank constant, $G$ is the gravitational constant, k is a constant, and $c$ is the speed of light.

Considering GUP in the framework of quantum geometry theory, any accelerating particle in the absence of gravity experiences a gravitational field [8]. This field corresponds to a perturbation unrelated to classical gravitation over the background Minkowski metric. This approach recovers GUP in the $p$-quadratic form as a function of the particle mass, $m$, the proper acceleration, $A$, and the quadratic form of the space-time length element, $\delta s$ [8]:

$$\Delta x \Delta p \geq \frac{\hbar}{2} + \frac{\hbar c^2}{m^2 A^2 \delta s^2} \Delta p^2 \, . \qquad (2)$$

However, the necessity for a minimum space–time length in quantum gravity theories still clashes with relativity, because this fundamental line element must be Lorentz-invariant. Hence, adapting the uncertainty principle to Lorentz covariance requires corrections over the GUP canonical momentum-position commutator in Minkowski space, described in [4,12]:

$$[P^{\mu}, X^{\nu}] = -i\hbar\left(1 + (\varepsilon - \alpha)\lambda^2 P^{\rho} P_{\rho}\right)\eta^{\mu\nu} - i\hbar(\beta + 2\xi)\lambda^2 P^{\mu} P^{\nu}. \qquad (3)$$

where the indices denoted by Greek letters take on the values 0 (time), 1, 2, and 3 (space) following standard tensor notation; In the context of this equation, $\alpha$, $\beta$, $\varepsilon$ and $\xi$ are dimensionless parameters to be adjusted to the specific problem, and $\lambda$ a parameter with dimensions of inverse momentum.

The uncertainty principle has also been used to tackle relativistic singularities. Some GR solutions contain singularities with infinite space–time curvature, for example within black holes [13-15]. However, a minimum allowable space–time line element would be incompatible with a point singularity. Thus, assuming quantized gravity and space–time, the uncertainty principle has been proposed to be the source of a repulsion force that prevents particles from reaching the singularity. The matter repelled from it would form a "Planck star" [16] or adopt a "fuzzball" structure [17]. This is important, because although infinite curvature is not problematic as a geometry, a black hole singularity could erase the history of any particle that ends up in it. This irreversible process would contribute to the black hole information paradox [18,19]. Nevertheless, it could be argued that particles in a black hole singularity could be distinguished by their proper time. However, the space–time line element is not defined right at the singularity. This is evident at radial coordinate position 0 in the Schwarzschild solution in spherical coordinates:

$$d\tau^2 = -\left(\frac{R - R_s}{R}\right)dt^2 + \frac{R \, dR^2}{R - R_s} + R^2 d\theta^2 + R^2 \sin^2\theta d\varphi^2. \qquad (4)$$

where $\boldsymbol{d\tau}$ corresponds to the proper time interval; $R$, $\theta$ and $\varphi$ stand for radial, polar and azimuthal coordinates in a Lorentzian metric, $g_{\mu\nu}$, signature (- + + +), and $R_s$ refers to the Schwarzschild radius. Towards the singularity, the time component of the metric, $g_{00}$, diverges to infinity while the radial component, $g_{RR}$, becomes 0. Additionally, the signs of these metric components change in the interior of the black hole once the event horizon is crossed, with the radial-dependent metric behaving as time rather than space. Despite these considerations, the singularity remains at $R$ position 0 where the length of the space–time line element is undefined.

a

In this paper, the classical principle of uncertainty was reformulated in terms of geometric parameters to impose a mathematical constraint on the geodesic equation. The requirement for a minimum space–time line element was confirmed in this re-formulation, as well as the need for imposing a perturbation over the background Minkowski metric unrelated to classical gravity. When applied to geodesics in the Schwarzschild metric, the presence of an exclusion zone around the singularity was also confirmed.

## 2. Derivation of a Relativistic Tensor Expression for the Classical Uncertainty Principle Inequalities

For simplicity, $c$ and the mass of the particle, m, are set to 1. Tensor notation is used throughout the paper. Contravariant coordinates are thus represented as $X^\mu$. In some specific cases, the temporal coordinate $X^0$ are represented as $t$ for clarity.

The classical uncertainty principle consists of two inequalities:

$$|\Delta p||\Delta x| \geq \frac{\hbar}{2} \quad , \quad |\Delta E||\Delta x^0| \geq \frac{\hbar}{2} \, . \tag{5}$$

These two inequalities can be written in tensor notation, following these identities:

$$|\Delta p||\Delta x| = \left|\sqrt{\Delta p^m \Delta p_m \, \Delta x_m \Delta x^m}\right| = |\Delta p_m \Delta x^m|$$

$$\Delta E \equiv \Delta P_0 \; ; \; |\Delta E||\Delta x^0| = \left|\sqrt{\Delta P^0 \Delta P_0 \, \Delta x_0 \Delta x^0}\right| = |\Delta P_0 \Delta x^0| \, . \tag{6}$$

In units of $c$ set to 1, energy is identified with the temporal component, $P^0$, of the relativistic 4-momentum vector ($P^\mu$) which is parametrized by proper time "$\tau$". Inequalities (5) take the following form in tensor notation:

$$|\Delta p_m \Delta X^m| \geq \frac{\hbar}{2} \quad ; \; |\Delta P_0 \Delta X^0| \geq \frac{\hbar}{2} \, . \tag{7}$$

The indices, denoted by Latin letters take on the values 1, 2 and 3 defining the spatial components.

To parametrize non-relativistic momentum by proper time, the gamma factor ($\gamma$) is introduced below, which is equivalent to the ratio between total energy (E) and mass energy (m) in units of c set to 1. Then the energy-time inequality is also modified with the gamma factor so that both can be represented by a single expression:

$$\left|\frac{1}{\gamma}\Delta P_m \Delta X^m\right| \geq \frac{\hbar}{2} \quad , \quad \left|\frac{1}{\gamma}\Delta P_0 \Delta X^0\right| \geq \frac{\hbar}{2\gamma} \quad , \quad \gamma = \frac{dt}{d\tau} \equiv \frac{E}{m} \, . \tag{8}$$

Adding Inequalities (8) one obtains:

$$|\Delta P_m \Delta X^m| + |\Delta P_0 \Delta X^0| \geq (1+\gamma)\frac{\hbar}{2} \, . \tag{9}$$

For non-relativistic particles, $\gamma$ is generally 1, recovering the expected classical expression:

$$|\Delta P_m \Delta X^m| + |\Delta P_0 \Delta X^0| \geq \hbar \, . \tag{10}$$

a

Assuming that the uncertainty principle must also apply to differential changes in momentum and position, the merged inequality can be rewritten:

$$|dP^m dX_m| + |dP_0 dX^0| \geq \frac{\hbar}{2} . \tag{11}$$

For simplification, the relativistic correction term $(1 + \gamma)$ in Inequality (9) is removed as it can be easily incorporated if required. Then, the inequality is expressed in terms of Planck length squared , $\ell_p^2$:

$$|dP^m dX_m| + |dP_0 dX^0| \geq \frac{\ell_p^2}{2G} . \tag{12}$$

The changes in the momentum and position are then expressed as rates of change with proper time:

$$\left| \frac{dP^m}{d\tau} d\tau \frac{dX_m}{d\tau} d\tau \right| + \left| \frac{dP^0}{d\tau} d\tau \frac{dX_0}{d\tau} d\tau \right| \geq \frac{\ell_p^2}{2G} , \tag{13}$$

which leads to:

$$\left| \frac{dP^m}{d\tau} U_m d\tau^2 \right| + \left| \frac{dP^0}{d\tau} U_0 d\tau^2 \right| \geq \frac{\ell_p^2}{2G} . \tag{14}$$

The uncertainty principle is thus reformulated in terms of the proper space–time length, 4-momentum change and proper velocity, $U_\mu$. It has to be noted that the inequality is undefined for null space–time length.

## 3. Derivation of a Covariant Geometric Form of the Uncertainty Principle

The reformulated Inequality (14) can be imposed over the geodesic equation. Geodesic trajectories can be identified with the rate of change of proper momentum in an interval of space–time length element:

$$\frac{dU^\mu}{d\tau} = -\Gamma^\mu_{\alpha\beta} U^\alpha U^\beta \equiv \frac{dP^\mu}{d\tau} , \tag{15}$$

where $\Gamma^\mu_{\alpha\beta}$ are Christoffel symbols calculated via the pseudo-Riemannian space–time metric tensor $g_{\mu\nu}$:

$$\Gamma^\mu_{\alpha\beta} = \frac{1}{2} g^{\mu\lambda} \left( \frac{\partial g_{\lambda\alpha}}{\partial x^\beta} + \frac{\partial g_{\beta\lambda}}{\partial x^\alpha} - \frac{\partial g_{\alpha\beta}}{\partial x^\lambda} \right) . \tag{16}$$

The geodesic equation (15) is incorporated in Inequality (14):

$$\left| U_m \Gamma^m_{\alpha\beta} U^\alpha U^\beta d\tau^2 \right| + \left| U_0 \Gamma^0_{\alpha\beta} U^\alpha U^\beta d\tau^2 \right| \geq \frac{\ell_p^2}{2G} . \tag{17}$$

Thus, the uncertainty principle is re-defined as the product of the interval of the space–time line element with a scalar derived from the geodesic trajectory (geometric scalar, $G_{geo}$):

$$G_{geo} \equiv 2G \left| U_m \Gamma^{\ m}_{\alpha\beta} U^\alpha U^\beta \right| + 2G \left| U_0 \Gamma^{\ 0}_{\alpha\beta} U^\alpha U^\beta \right| ,$$

$$\left| G_{geo} \, d\tau^2 \right| \geq \ell_p^2 . \tag{18}$$

This form of the uncertainty principle imposes a minimum line element for proper space–time distance related to Planck length through a Lorentz-covariant geodesic scalar. This inequality also estimates a degree of uncertainty in the geodesic trajectory.

## 4. Geometric Uncertainty Principle in Minkowski space

Here, GeUP is applied for a particle at rest in classical Minkowski space with a (- + + +) Lorentz metric signature and the Minkowski metric tensor denoted as $\boldsymbol{\eta}_{\mu\nu}$. This metric has null Christoffel connectors, resulting in the geodesic scalar being 0. Then, Inequality (18) represents a contradiction unless Planck length is considered 0 in the non-quantum limit:

$$0 \geq \ell_p^2 . \tag{19}$$

To comply with GeUP, the Minkowski metric has to deviate from flat space, for example by introducing a differential perturbation, $h_{\mu\nu}$. The resulting metric tensor $g_{\mu\nu}$ then reads:

$$g_{\mu\nu} = \eta_{\mu\nu} + h_{\mu\nu} . \tag{20}$$

In this example, the perturbation depends only on the temporal coordinate to fulfil conditions of spatial homogeneity and isometry. For a particle at rest, only the temporal component of the proper velocity is non-zero. Then the only relevant component of the metric tensor for the calculations is:

$$g_{00} = \eta_{00} + h_{00} = -1 - \varepsilon(t) . \tag{21}$$

Here, the function $\varepsilon(t)$ is defined as the time-dependent perturbation corresponding to $h_{00}$, to avoid confusion with the Planck constant. For a particle at rest, only the $X^0$ coordinate ($t$) contributes to proper time. The Inequality (18) takes the following form in terms of the Planck constant:

$$\left| -2U_0 \Gamma^{\ 0}_{00} U^0 U^0 \, (-1-\varepsilon) \, dt^2 \right| \geq \hbar . \tag{22}$$

The calculation of the Christoffel symbol is straightforward because only $g_{00}$ has a dependency in the $X^0$ coordinate (Equation 21):

$$\Gamma^{\ 0}_{00} = \frac{1}{2} g^{00}(\partial_0 g_{00}) = \frac{-1}{2(1+\varepsilon)} \partial_0(-1-\varepsilon) = \frac{\partial_0 \varepsilon}{2(1+\varepsilon)} ,$$

$$\partial_0 = \frac{\partial}{\partial x^0} \equiv \frac{\partial}{\partial t} . \tag{23}$$

Using the term (23), Inequality (22) can be solved:

$$\left| -2U_0 U^0 U^0 \, \frac{\partial_0 \varepsilon}{2(1+\varepsilon)} \, (-1-\varepsilon) \, dt^2 \right| \geq \hbar ,$$

$$|-U^0\,\partial_0\varepsilon\ \ dt^2|\geq\hbar\ .\tag{24}$$

This inequality can be simplified by multiplying $\partial_0\varepsilon$ by $dt$ and reintroducing the mass of the particle. This last step replaces $U^0$ by the temporal component of the 4-momentum vector:

$$|P^0\,d\,\varepsilon\ \ dt\,|\geq\hbar\ .\tag{25}$$

The classical expression for time-energy uncertainty is recovered. Thus, one can determine the energy of the particle in a given interval of geodesic time $dt$ to be $P^0$, with $d\varepsilon$ denoting the accuracy on the measurement. High-precision determination of $P^0 d\varepsilon$ implies long intervals of time. Likewise, measurements over increasingly precise intervals of geodesic time correspond to increased fluctuations $(d\varepsilon)$ in the energy of the particle. These fluctuations of $\varepsilon$ alter the background space–time metric (Equation 21). The relativistic factor omitted in Inequality (9) can also be re-introduced with a value of 2 leading then to:

$$\left|P^0\ d\,\varepsilon\ \ dt\,\right|\geq 2\hbar\quad,\qquad |d\varepsilon|\geq\left|\frac{2\hbar}{P^0\,dt}\right|\ .\tag{26}$$

The inequality can be changed to an equation for the minimum allowable value. To illustrate, the differential $d\varepsilon(t)$ is approximated to an interval:

$$\varepsilon=\varepsilon_0+\frac{2\hbar}{P^0\,dt}.\tag{27}$$

The initial value for the perturbation field can be chosen as 0 (no correction over the Minkowski metric). This leads to a minimally allowed perturbation of the metric as follows:

$$g_{00}=-1-\frac{2\hbar}{P^0 dt}\ .\tag{28}$$

In this case, the metric perturbation over the Minkowski metric background is unrelated to classical gravitation. It is thus a function of the accuracy in the interval of coordinate time in the geodesic trajectory. Minkowski space–time is recovered in the non-quantum limit, $\hbar\to 0$.

## 5. Application of the Geometric Uncertainty Principle to the Schwarzschild Metric

The Schwarzschild metric corresponds to the solution for a point mass generating a gravitational field. This solution has spherical symmetry and is usually expressed in spherical coordinates (Equation 4). The spherical symmetry allows the selection of physically relevant trajectories in space–time geodesics with constant polar and azimuthal coordinates. These conditions ensure the following statements:

$$d\theta=d\varphi=0\ \ \to\ \ U^\theta=U^\varphi=0\quad,\ \ U_0U^0+U_RU^R=-1.\tag{29}$$

These geodesics have contributions from $t$ and $R$ components of proper velocity. Inequality (18) then reads:

$$\left|dR^2+\frac{g_{00}}{g_{RR}}dt^2\right|\geq\left|\frac{\ell_p^2}{g_{RR}G_{\text{geo}}}\right|\ .\tag{30}$$

Introducing the components of the metric tensor, one finds:

a

$$\left| dR^2 + \frac{(R - R_s)^2}{R^2} dt^2 \right| \geq \left| \frac{(R - R_s)\, \ell_p^2}{R\, G_{\text{geo}}} \right| . \tag{31}$$

The geodesic scalar is calculated with the contributions from temporal and radial components of proper velocity:

$$G_{\text{geo}} = 2G \left| U_0 \Gamma^{\,0}_{\alpha\beta} U^\alpha U^\beta \right| + 2G \left| U_R \Gamma^{\,R}_{\alpha\beta} U^\alpha U^\beta \right| . \tag{32}$$

The non-zero Christoffel connectors relevant for this solution are as follows:

$$\Gamma^{\,0}_{R0} = -\Gamma^{\,R}_{RR} = \frac{R_s}{2R(R - R_s)} \quad , \quad \Gamma^{\,R}_{00} = \frac{R_s(R - R_s)}{2R^3} \quad , \tag{33}$$

leaving the geometric scalar as:

$$\begin{aligned} G_{\text{geo}} = 2G \big( & \left| U_0 \Gamma^{\,0}_{R0} U^R U^0 + U_0 \Gamma^{\,0}_{0R} U^0 U^R \right| \\ & + \left| U_R \Gamma^{\,R}_{00} U^0 U^0 + U_R \Gamma^{\,R}_{RR} U^R U^R \right| \big) . \end{aligned} \tag{34}$$

This expression is solved as:

$$G_{\text{geo}} = \frac{G R_s U^R (2 U_0 U^0 + 1)}{R(R - R_s)} . \tag{35}$$

Inserting Equation (35) into Inequality (31) and solving for the square of the length element in the $R$ coordinate:

$$dR^2 \geq \left| \frac{2M}{(2 U_0 U^0 + 1) U^R} \left( \frac{R}{R_s} - 1 \right)^2 \ell_p^2 - \left( 1 - \frac{R_s}{R} \right)^2 dt^2 \right| \tag{36}$$

The gravitational constant $G$ is replaced by $R_s/2M$ with $M$ representing the mass generating the gravitational field. In geodesics close to the singularity, the uncertainty in the $R$ coordinate (represented by $dR^2$) diverges to infinity. Particles close to the singularity are thus highly de-localized. No particle geodesic below the threshold, set by Inequality (36), are allowed. This condition defines an exclusion zone around the singularity as a function of uncertainty in the $t$-coordinate, see Figure 1.

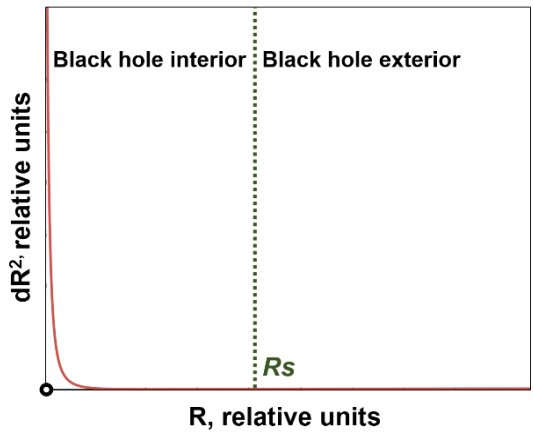

**Figure 1.** Black hole singularity exclusion zone. Relative uncertainty, $dR^2$, in the radial ($R$) coordinate as a function of $R$ (Equation (36)). The uncertainty diverges to infinity towards the singularity

($R = 0$). Geodesics with $dR^2$ values below the curve are not allowed, and define a particle exclusion zone in the interior of the black hole.

Let us now consider the case when the $t$-coordinate of a particle in the geodesics, defined by Equation (36), is known with absolute certainty. It has to be stressed that this condition leads to the minimum possible uncertainty in the R coordinate when approaching the singularity. This implies:

$$dt = 0 \,. \tag{37}$$

Additionally, the solution to inequality (36) is:

$$dR^2 \geq \left| \frac{2M}{U^R} \left( \frac{R}{R_s} - 1 \right)^2 \ell_p^2 \right| \,. \tag{38}$$

Thus, the uncertainty in $R$ reaches a well-defined, minimum allowed value right in the singularity at $R = 0$, which corresponds to:

$$dR^2(R = 0) \geq \frac{2M}{U^R} \ell_p^2 \,. \tag{39}$$

By selecting geometries with constant $t$ coordinate, a threshold value for uncertainty in $R$ can be calculated, below which no geodesic is ever allowed. An approximate calculation of $dR$ for a stellar mass black hole gives a value within the range of $10^{-15}$ to $10^{-16}$ cm.

## 5. Discussion

In this paper, the principle of uncertainty is reformulated in terms of covariant geometric parameters. This reformulation is the result of applying the classical uncertainty principle inequalities over the geodesic equation just as a mathematical constraint. No quantization of space–time or quantum gravitational field are introduced. Hence, it has to be remarked that the current study neither constitutes a theory for quantum gravity, nor represents a canonical generalization on quantum gravity. However, the uncertainty inequalities presented in this form yield interesting results that can be interpreted in light of current quantum gravity theories.

First, there is an imposition of a minimum length for the space–time line element. This is a common feature of current quantum gravity theories such as loop quantum gravity [20,21], string theory [10,19,22,23] and doubly special relativity [24]. Therefore, although a geometric expression for the uncertainty principle (GeUP) is not a quantum gravity theory per se, it introduces ambiguity in space–time trajectories. The space–time distance element is expressed as a relationship with Planck length through a geodesic-derived scalar, which ensures Lorentz covariance.

Second, in agreement with results from a generalized uncertainty principle (GUP) [8,11], GeUP is shown here to be incompatible with Minkowski space. Its mathematical constraint over the geodesics imposes a perturbation in the metric unrelated to classical gravity. This incompatibility is to be expected in any quantum gravity theory because vacuum energy fluctuations are likely to alter the metric [8]. In the example, described in this paper, the perturbation is a function of the interval of coordinate time in the geodesic trajectory, and the energy of the particle. GUP and GeUP enforce a perturbation in the metric of Minkowski space which can be considered a part of a gravitational background state. This perturbation alters the classical uncertainty principle in GUP by a factor dependent on the mass-energy and proper acceleration (see Inequality (2)) [8].

GUP has been applied to multiple scenarios such as black-hole entropy, and thermodynamics of cosmological models, as it is extensively reviewed in [25]. Even so, corrections over its canonical expression are also required to ensure Lorentz invariance [12]. It could be argued that metric perturbations caused by the uncertainty principle constitute a background state for the gravitational field. However, the concept for gravitational background states in general relativity (GR) is highly problematic because classical GR is a non-linear but otherwise background-independent theory [9]. Nevertheless, vacuum energy and dark energy may contribute to potential gravitational background states and to the expansion of the universe through the cosmological constant $\Lambda$ [26]. So far, the calculations on the contributions to $\Lambda$ by different sources do not provide a satisfactory explanation to its small positive value [27].

Third, the uncertainty principle has been proposed to be the source of a repulsion force from black hole singularities. This "uncertainty force" is the basis for the stability of Planck stars [16]. Application of GeUP over the Schwarzschild solution also uncovered a region close to the singularity below which geodesics violate the uncertainty principle. Particles approaching the singularity are expected to have an uncertainty in the radial ($R$) coordinate so large that these particles would appear to be repelled from the singularity. Here, the minimum allowed uncertainty ($dR$) in the $R$ coordinate is calculated for a stellar mass black hole to be within the order of $10^{-15}$ cm. A rough estimation on the diameter of a Planck star by loop quantum gravity gives a value of about $10^{-10}$ cm [16,20]. Both calculations provide sizes several orders of magnitude larger than Planck length.

Concluding, in this paper, the classical uncertainty principle is re-formulated to impose a mathematical restriction on GR geodesics. The equations obtained do not represent a bona fide quantum gravity theory but highlight some of the common issues to quantum gravity theories: a minimum length for the space–time line element that must keep Lorentz invariance, and the need for background quantum gravity states. However, complete answers to the several drawbacks have to be constructed from well-founded quantum gravity theories based on canonical principles.

**Author Contributions:** Conceptualization, D.E. and G.K.; methodology, D.E; resources, D.E. and G.K. All authors have read and agreed to the published version of the manuscript.

**Funding:** D.E. is funded by a Miguel Servet Fellowship (ISCIII, Spain, Ref CP12/03114).

**Conflicts of Interest:** The authors declare no conflict of interest.